\documentclass[prd,preprintnumbers,nofootinbib,superscriptaddress,onecolumn,notitlepage]
{revtex4}

\usepackage{cancel}
\usepackage{amsmath,mathrsfs,amscd,graphicx}
\usepackage{mciteplus}
\usepackage{multirow}
\usepackage{color}
\usepackage{cancel}
\usepackage{rotating}
\usepackage{slashed}
\usepackage{subcaption}
\usepackage{textcomp}
\usepackage[final]{pdfpages}
\usepackage{array}
\usepackage{float}
\usepackage{url}
\usepackage{textcomp}
\usepackage{amsfonts, latexsym, epsfig}
\usepackage{bm}
\usepackage{times}
\usepackage{epsfig}
\usepackage{amssymb}
\usepackage{tikz}
\usepackage{slashed}
\usepackage{hyperref}
\usepackage{verbatim}
\usepackage[normalem]{ulem}
\usepackage[utf8]{inputenc}
\usepackage{diagbox}
\usepackage{titlesec}
\def\beq{\begin{equation}}
\def\eeq{\end{equation}}
\def\be{\begin{equation}}
\def\ee{\end{equation}}
\def\bea{\begin{eqnarray}}
\def\eea{\end{eqnarray}}

\definecolor{dpmagenta}{rgb}{0.8, 0.0, 0.8}

\allowdisplaybreaks[3]
\begin{document}
\title{Propagation of angular momentum in charged pion decay and related processes}
\author{Bowen Wang~}
\email{bowenw@hznu.edu.cn}
\affiliation{School of Physics, Hangzhou Normal University, Hangzhou, Zhejiang 311121, China}
%

\begin{abstract}

	There are confusions about angular momentum propagation in scattering or decay processes involving the transition between particle systems
	that appear to transform differently under Lorentz transformations. This paper provides an analysis of the transformation 
	properties of the states and interactions for a few typical processes within the standard model of particle physics, 
	and performs explicit calculations showing how angular momentum transfers in these processes. We shall show explicitly 
	a) how a state with zero angular momentum evolves via interactions mediated by a single vector boson of spin one, and b) that 
	angular momentum conservation is completely consistent with the	calculation in quantum field theory. A discussion is also given 
	about the phenomenological consequences of the theoretical results obtained in this study.
	
\end{abstract}

\maketitle

\section{Introduction}

\begin{figure}[h]
\centering
    \includegraphics[scale=0.5]{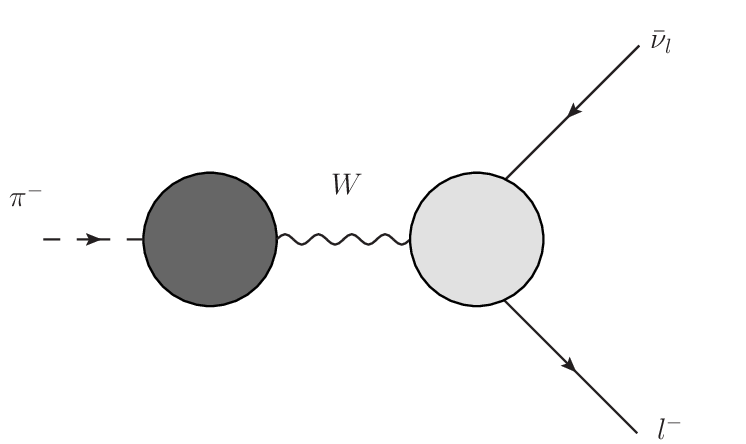} \qquad
	\caption{The decay of a charged pion into a lepton and an anti-neutrino: $\pi^-\rightarrow l^- \bar{\nu}_l$. 
	The light blob represents purely perturbative interactions in the graph, while the dark blob also includes non-perturbative contributions.}
\label{fig:pidec}
\end{figure}

Within the standard model (SM), the decay of a charged pion\footnote{Or more generally, 
a charged pseudoscalar meson.} (take $\pi^-$ to be definite) into a lepton $l^-$ 
and an anti-neutrino $\bar{\nu}_l$ occurs through weak interactions mediated by a W boson. 
This process is described in Fig.~\ref{fig:pidec}. 
The corresponding matrix element is 
\begin{align}
	&	(2\pi)^4\delta^{(4)}(k-k_1-k_2)i\mathcal{M}(\pi^-(k)\rightarrow l^-(k_1) \bar{\nu}_l(k_2)) \nonumber \\
	&=-\frac{i\sqrt{2\pi\alpha} V_{ud}}{\sin\theta_W}\int d^4x\langle l^-(k_1) \bar{\nu}_l(k_2)|W^+_{\mu}(x)\ |0\rangle \times \langle 0|J_q^{\mu +}(x)|\pi^-(k) \rangle,
\label{eq:pidec} 
\end{align}
where $\alpha$ is the fine structure constant, $V_{ud}$ the ``$ud$'' element of the quark mixing matrix, and $\theta_W$ the weak mixing angle.
The definition of the amplitude $\mathcal{M}$ follows the convention of Ref.~\cite{Peskin:1995ev}.
The pion and leptons are in ``in'' and ``out'' states respectively, and the fields between the states are all renormalized.
The polarizations of the fermions\footnote{We do not distinguish anti-fermions from fermions except where the distinction affects our discussion.} 
are not labeled explicitly. The $V-A$ quark current is defined by
\begin{gather}
	J_q^{\mu +}(x)\equiv \overline{\Psi^u}(x)\gamma^{\mu}\frac{1-\gamma^5}{2}\Psi^d(x).
\label{eq:VAcurrents} 
\end{gather}
Note that the decay occurs inevitably via the interaction term $W^+_{\mu}(x)J_q^{\mu +}(x)$, which is explicitly extracted in Eq.~\ref{eq:pidec}.
The insertion of the vacuum states indicates then that the fields can only be contracted in such a way as to maintain 
a factorized form for the transition amplitude represented by the graph in Fig.~\ref{fig:pidec}.
The first factor of the integrand in Eq.~\ref{eq:pidec} gives the vertex function of two external fermions
and an off-shell propagator~\cite{Weinberg:1995mt,Sterman:1993hfp} with all order corrections including the self-energy graphs.
\footnote{Only half of the
self-energy contribution should be included for the external lines 
in accord with the reduction of the S-matrix~\cite{Weinberg:1995mt,Sterman:1993hfp}.}  
These are denoted by the blob in light grey in Fig.~\ref{fig:pidec}. The second factor has a simple tensor structure determined 
by the properties of the quark current and of the states under Lorentz transformations\footnote{The Lorentz 
transformations (and Lorentz group) in this paper do not involve space inversion and time reversal operations.} and translations:
\begin{gather}
	\langle 0|J_q^{\mu +}(x)|\pi^-(k) \rangle = \frac{i}{2}f_{\pi}k^{\mu}e^{-ik\cdot x},
\label{eq:piVert} 
\end{gather}
where $f_{\pi}$ is the pion decay constant (in the convention of the particle data group~\cite{ParticleDataGroup:2022pth}). 
It receives contributions from perturbative and non-perturbative interactions
within the $\pi^-$ induced by the quark current $J_q^{\mu +}(x)$. These interactions result in 
the decay of the pion and are represented by the blob in dark grey in Fig.~\ref{fig:pidec}.

Perturbative calculations can be done for the branching ratio 
$R_{e/\mu}^{\pi}\equiv \Gamma(\pi^-\rightarrow e^- \bar{\nu}_e)/\Gamma(\pi^-\rightarrow \mu^- \bar{\nu}_{\mu})$,
which is independent of $f_{\pi}$. The agreement between the SM prediction of $R_{e/\mu}^{\pi}$ (with radiative corrections) 
and measurements has reached a level within 0.5\%, and is limited by the accuracy of the experiment~\cite{Pocanic:2014jka}.
This provides a good precision test of the SM and perturbative calculations in quantum field theory. Furthermore, $f_{\pi}$
can be calculated in lattice QCD (see Ref.~\cite{ParticleDataGroup:2022pth}, Table 72.1). 
It can also be extracted from experiments once the W decay subprocess 
(i.e., the first factor of the integrand on the r.h.s. of Eq.~\ref{eq:pidec}) 
is computed (with parameters such as $V_{ud}$ taken from other measurements). 
Comparison of results from both approaches shows an agreement at the percent level~\cite{ParticleDataGroup:2022pth}. 
This is also the level of uncertainty of the Lattice QCD result.

On the other hand, measurements of the angular distributions of pion decay products in the pion rest frame were done in many
early experiments. While most of them obtained isotropic results (e.g., Ref.~\cite{Abashian:1957zz,PhysRevLett.14.745,Frota-Pessoa:1969jeg}),  
small asymmetry were found in some of the studies (e.g., Ref.~\cite{Hulubei:1963zza} ).  
There are no reports of significant deviation from a uniform angular distribution 
in more recent experiments of pion decays~\cite{MINOS:2008fnv,MINOS:2012ozn}. The limits for the sizes of possible Lorentz violating terms are set to
the level of $10^{-5}\sim 10^{-4}$ relative to the SM terms~\cite{Altschul:2013yja,Altschul:2013ykb,Noordmans:2014bua}.

It is a consensus that no anisotropic angular distributions can be deduced from the charged pion's SM decays, 
since the pion is a pseudoscalar and the interactions involved are all Lorentz invariant.
Studies of possible Lorentz violating effects all seek to find the sources for Lorentz violation from 
extensions of the SM~\cite{Altschul:2013yja,Altschul:2013ykb,Diaz:2013saa,Noordmans:2013dha,Noordmans:2014bua}.
It may be conceptually confusing, however, that in the SM a spin-0 initial state propagates via a spin-1 vector boson, 
and remains isotropic before being measured. This is usually explained by noting that there is a longitudinal polarization mode 
of the massive vector boson that behaves like a scalar with 0 angular momentum~\cite{Fayyazuddin:1994wh,Peskin:1995ev,Barger:1987nn}.
In Ref.~\cite{Nakanishi:2002sv} the decay is considered as occuring via a spin-0 goldstone boson to avoid non-conservation of angular momentum.
Nevertheless, it is still helpful to analyze the decay and other related processes using symmetry principles to show explicitly how angular momentum 
conservation manifests at the amplitude level.

In the following we shall first explore the rotational and Lorentz symmetries in production processes of a vector boson, 
and then illustrate the results based on symmetry considerations by simple tree level calculations. Consequences of the results in
experiments will then be discussed. A summary and discussion of the results will be given at the end.

\section{Angular momentum transfer in the production of a vector boson}
\label{sec:gen_disc}

\begin{figure}[!ht]
  \centering
	\begin{subfigure}[t!]{0.3\linewidth}
	  \parbox[][4cm][c]{\linewidth}{
      \centering
	  \includegraphics[scale=0.5]{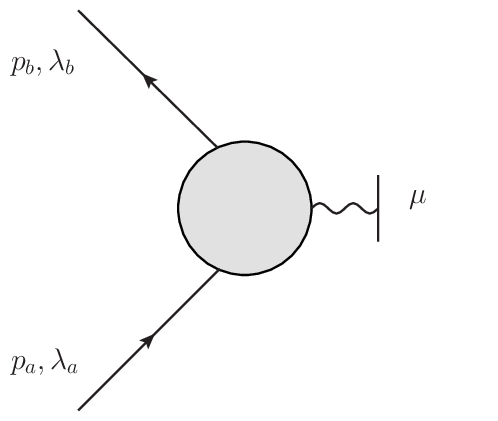}
	}
	  \subcaption{}
  \end{subfigure}
  \begin{subfigure}[t!]{0.3\linewidth}
	  \parbox[][4cm][c]{\linewidth}{
      \centering
	  \includegraphics[scale=0.5]{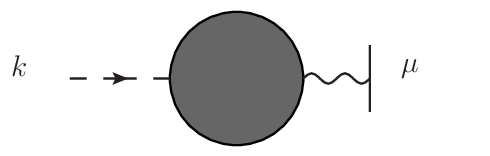}
	  }
	  \subcaption{}
  \end{subfigure}
	\caption{Vertices that produce a vector boson in the processes (a): fermion-fermion scattering, and (b): decay (of e.g., $\pi^{\pm}$, $\Psi$, etc.).
	The light blob represents purely perturbative interactions in the graph, while the dark blob also includes non-perturbative contributions.}
\label{fig:Vert}
\end{figure}

It is helpful to first work out the angular momentum of a vector boson (vector current) created from a collision of two spin-$1/2$ fermions. 
This is somewhat more complicated than the case of charged pion decay, but it provides a richer picture as to how angular momentum is transferred
between different parts of an amplitude. To make the discussion general, we start by considering the matrix element
\begin{gather}
	\langle 0 | J^{\mu}(x) | p_a,\lambda_a; p_b, \lambda_b \rangle 
	= e^{-i(p_a+p_b)\cdot x}\langle 0 | J^{\mu}(0) | p_a,\lambda_a; p_b, \lambda_b \rangle,
\label{eq:Vert}
\end{gather}
where $J^{\mu}$ denotes the SM current that tranforms like a vector, axial-vector, or a combination of them, 
e.g., $J^{\mu} =\overline{\Psi^e}\gamma^{\mu}\Psi^e$, $\overline{\Psi^e}\gamma^{\mu}\frac{1-\gamma^5}{2}\Psi^{\nu}$, etc.
The momenta and helicities of the fermions are indicated in the state ket.
This matrix element represents the all order contribution to the fermion-vector vertex  
shown in Fig.~\ref{fig:Vert} (a) (with the vector boson part excluded, as indicated by the vertical bar).

The initial state is made up of two fermions with definite momenta and helicities, as indicated in the state ket in Eq.~\ref{eq:Vert}. This is
not an eigenstate of the angular momentum. 
To analyse the angular structure of the process, we expand the initial state as a combination of states with 
definite  $J$, $M$, $\lambda_a$, $\lambda_b$, $\bold{P}^+\equiv \bold{p}_a+\bold{p}_b$, 
and $E^+ \equiv  \sqrt{\bold{p}_a^2+m_a^2}+\sqrt{\bold{p}_b^2+m_b^2}$~\cite{Jacob:1959at,Gibson:1976wp}.
$J$ and $M$ are respectively the quantum numbers for the total angular momentum of the fermions, and for its projection on the 3-axis in
the center-of-mass (CM) frame of the fermions, in which the 3-direction is defined by the direction of the 3-momentum of a: 
$\bold{p}\equiv \bold{p}_a =-\bold{p}_b$. It is convenient to work in this CM frame and express the expansion as~\cite{Jacob:1959at,Gibson:1976wp}
\begin{gather}
	| p_a,\lambda_a; p_b, \lambda_b \rangle = \sum_J C_J |J, M=\lambda_a-\lambda_b, \lambda_a, \lambda_b, |\bold{p}|\rangle,
\label{eq:stateexp}	
\end{gather}
where $\bold{P}^+=0$ is suppressed, and $E^+$ is replaced by $|\bold{p}|$ in the states on the r.h.s. of Eq.~\ref{eq:stateexp}.
$J$ runs over the integers $0, 1, \ldots$. There is no sum over $M$, because the 3-component of the angular momentum 
coincides with the difference of the helicities. 
For an anti-fermion, the physical spin in the direction of its momentum is opposite to its helicity, so
an extra minus sign\footnote{This extra minus sign is not to be confused with the minus sign in $M=\lambda_a-\lambda_b$.} 
is to be included in its helicity $\lambda_a$ or $\lambda_b$ when obtaining $M=\lambda_a-\lambda_b$.
It is shown~\cite{Jacob:1959at,Gibson:1976wp} that $C_J=\sqrt{(2J+1)/(4\pi)}$ in order to be consistent with 
the normalization of the momentum and angular momentum eigenstates:
\begin{gather}
	\langle  p_a^{\prime},\lambda_a^{\prime}; p_b^{\prime}, \lambda_b^{\prime} | p_a,\lambda_a; p_b, \lambda_b \rangle = (2\pi)^6 2E_a 2E_b \delta^3(\bold{p}_a^{\prime}-\bold{p}_a)\delta^3(\bold{p}_b^{\prime}-\bold{p}_b)\delta_{\lambda_a^{\prime} \lambda_a}\delta_{\lambda_b^{\prime} \lambda_b},
\label{eq:momstatesnorm} \\
	\langle  J^{\prime},M^{\prime},\lambda_a^{\prime},\lambda_b^{\prime}, |\bold{p}^{\prime}|| J,M, \lambda_a, \lambda_b, |\bold{p}| \rangle = (2\pi)^6\frac{E^+}{|\bold{p}|}\delta^3(\bold{P}^{+\prime})\delta(E^{+\prime}-E^+)\delta_{J^{\prime}J}\delta_{M^{\prime}M}\delta_{\lambda_a^{\prime} \lambda_a}\delta_{\lambda_b^{\prime} \lambda_b}.
\label{eq:Angstatesnorm} 
\end{gather}
Note that Eq.~\ref{eq:Angstatesnorm} is expressed in the CM frame of $\bold{P}^+$, in which $\bold{P}^{+\prime}$ may be non-zero. 
For a non-zero $\bold{P}^{+\prime}$, Eq.~\ref{eq:Angstatesnorm} gives 0 irrespective of the 3-direction chosen in the CM frame of $\bold{P}^{+\prime}$
to define $M^{\prime}$. Whereas if $\bold{P}^{+\prime}=\bold{P}^{+}=0$, both $M$ and $M^{\prime}$ are defined with 
respect to the direction of $\bold{p}$. However, in Eq.~\ref{eq:Angstatesnorm}, $M$ and $M^{\prime}$ are not restricted to 
$M=\lambda_a-\lambda_b$ and $M^{\prime}=\lambda_a^{\prime}-\lambda_b^{\prime}$, since the states are not restricted to those in the expansion
in Eq.~\ref{eq:stateexp}.  The matrix element in Eq.~\ref{eq:Vert} can also be expanded as
\begin{gather}
	\langle 0 | J^{\mu}(0) | p_a,\lambda_a; p_b, \lambda_b \rangle 
	= \sum_J C_J\langle 0 | J^{\mu}(0)|J, M=\lambda_a-\lambda_b, \lambda_a, \lambda_b, |\bold{p}|\rangle,
\label{eq:matexp} 
\end{gather}
or, for a full process with the final state $| f \rangle$,
\begin{gather}
	\langle f | p_a,\lambda_a; p_b, \lambda_b \rangle = \sum_J C_J\langle f|J, M=\lambda_a-\lambda_b, \lambda_a, \lambda_b, |\bold{p}|\rangle.
\label{eq:matexp2} 
\end{gather}

We are particularly interested in analysing the contribution to the matrix element in Eq.~\ref{eq:matexp} from each value of $J$.  
A vanishing contribution from a specific $J$ signifies that the propagation of this angular momentum eigenvalue is forbidden.
To determine the allowed $J$ values we should consider the transformation properties of both the operator $J^{\mu}$ and the state 
$|J, M=\lambda_a-\lambda_b, \lambda_a, \lambda_b, |\bold{p}|\rangle$. The angular momentum eigenstates with fixed $J$
furnish an irreducible representation under the spatial rotation $R(\alpha, \beta, \gamma)$, which is a 3 by 3 matrix that rotates
the spatial coordinates and is parameterized by the Euler angles $\alpha, \beta$ and $\gamma$:
\begin{gather}
	U(R(\alpha, \beta, \gamma))	|J, \lambda_a-\lambda_b, \lambda_a, \lambda_b, |\bold{p}|\rangle = \sum_{M^{\prime}} \mathscr{D}^J_{M^{\prime}, \lambda_a-\lambda_b}(R(\alpha, \beta, \gamma)) |J, M^{\prime}, \lambda_a, \lambda_b, |\bold{p}|\rangle.
\label{eq:kettransf}
\end{gather}
Note that $\lambda_a$, $\lambda_b$, and  $|\bold{p}|$ are invariant under rotations. 
$M$ (on the l.h.s.) is fixed to $\lambda_a-\lambda_b$, 
but is changed by a rotation to $M^{\prime}$ (on the r.h.s.). 
 The subscripts of the Wigner $\mathscr{D}$-matrix $\mathscr{D}^j_{m^{\prime},m}$ (The explicit form 
 can be found in e.g., ~\cite{Jacob:1959at,Leader:2011vwq}) take the  values $-J, -J+1, \ldots, J$. 
 (All helicity values $\lambda_a$ and $\lambda_b$ are allowed as long as $-J\leq M=\lambda_a-\lambda_b\leq J$).
 The current operator $J^{\mu}(0)$ is a 4-vector that transforms according to 
\begin{gather}
	U(\Lambda) J^{\mu}(0)U^{-1}(\Lambda) = (\Lambda^{-1})^{\mu}_{\nu} J^{\nu}(0)
\label{eq:currtransf}
\end{gather}
under an arbitrary Lorentz transformation $\Lambda$ (The $x$ dependence of $J^{\mu}$ is removed by translation in Eq.~\ref{eq:Vert} so that
the transformation of the current is simpler).  On the r.h.s. of Eq.~\ref{eq:currtransf} $\mu$ and $\nu$ label respectively 
the row and column of the matrix $\Lambda^{-1}$. This is equivalent to the self-representation of the Lorentz group, which can be seen
from the property for an arbitrary Lorentz transformation matrix $\Lambda$ 
\begin{gather}
	g \Lambda g^{-1}= (\Lambda^{-1})^T, \\
	g=g^{-1}= \text{diag}\{1,-1,-1,-1\}.
\end{gather}
Therefore $J^{\mu}(0)$ furnish a 4-dimensional irreducible representation of the Lorentz group. However, upon restriction 
to its spatial rotation subgroup $SO(3)$, the current operators are no longer irreducible. Clearly,  $J^0$ is invariant under rotations, 
while $J^1$, $J^2$, and $J^3$ are mixed. This reduces the 4 components to an $SO(3)$ singlet labeled
by $J=0$ and a triplet in the $SO(3)$ self-representation labeled by $J=1$:
\begin{gather}
	U(R) J^0(0)U^{-1}(R) = J^0(0) D^{J=0}_{0,0}(R),  \quad D^{J=0}(R) = 1, \\
	U(R) J^i(0)U^{-1}(R) = J^{j}(0) D^{J=1}_{j,i}(R), \quad D^{J=1}_{j,i}(R) = (R^{-1})_{i,j}= R_{j,i}, \quad i, j = 1,2,3.
\end{gather}
As representations of the $SO(3)$ group with the same dimension are equivalent, 
the $D^{J=0}$ and $D^{J=1}$ matrices above are equivalent to the representations 
$\mathscr{D}^0$ and $\mathscr{D}^1$ in Eq.~\ref{eq:kettransf}, respectively.
 Hence the only non-vanishing contribution in Eq.~\ref{eq:matexp} is from $J=0$ and $J=1$ 
 according to Wigner-Eckart theorem.\footnote{The rule of adding two angular momenta dictates that $J^{\mu}(0)$ and 
$|J, M=\lambda-\lambda^{\prime}, \lambda, \lambda^{\prime}, |\bold{p}|\rangle$ must be in equivalent representations
 in order to be combined by Clebsch-Gordon coefficients to form a singlet under rotations.} 
 This is not exactly right because they could still vanish due to constraint from further 
 symmetries (See section~\ref{sec:ex_proc} for an example), but
 at least all final states with $J\geq 2$ cannot be produced via a vector boson coupling to the vertex in Fig.~\ref{fig:Vert} (a).
 
One may want to also investigate the transformation of $J^{\mu}(0)$ and 
$|J, M=\lambda-\lambda^{\prime}, \lambda, \lambda^{\prime}, |\bold{p}|\rangle$ 
under the full Lorentz group, rather than its subgroup. 
However, it is not straightforward to see the transformation law of the angular momentum eigenstates in this case, 
except that $|J=0, M=\lambda-\lambda^{\prime}=0, \lambda, \lambda^{\prime}, |\bold{p}|\rangle$ will no longer 
be invariant (a boost changes the eigenvalues $\bold{P}^+$ and $E^+$).

The analysis above can be carried over immediately to the case of the charged pion decay (or more generally, charged pseudoscalar meson decay) 
via production of a W boson, as in Fig.~\ref{fig:Vert} (b). 
The initial state pion is a pseudoscalar with $J=M=0$, and
the amplitude $\langle 0|J_q^{\mu +}(x)|\pi^-(k) \rangle$ in Eq.~\ref{eq:piVert} is non-zero from
the same reasoning as given above of the $J=0$ contribution being non-zero in two fermion scatterings.

\section{Amplitudes with angular momentum eigenstates}

Now we derive formulae for practical calculations of the amplitudes on the r.h.s. of Eq.~\ref{eq:matexp2}. 
The spatial part of $p_a$ in Eq.~\ref{eq:stateexp} can be rotated by $R(\phi, \theta, 0)$ to a direction determined
by the azimuthal angle $\phi$ and polar angle $\theta$, with respect to the original axes. Denoting the initial state under this rotation by
\begin{gather}
	|\phi, \theta, \lambda_a, \lambda_b, |\bold{p}|\rangle \equiv U(R(\phi, \theta, 0))| p_a,\lambda_a; p_b, \lambda_b \rangle
	=| R(\phi, \theta, 0)p_a,\lambda_a; R(\phi, \theta, 0)p_b, \lambda_b \rangle , \\
	| p_a,\lambda_a; p_b, \lambda_b \rangle = |\phi=0, \theta=0, \lambda_a, \lambda_b, |\bold{p}|\rangle,
\label{eq:rotstate}	 
\end{gather}
we obtain the expansion of the momentum eigenstates in an arbitrary direction by applying the rotation on Eq.~\ref{eq:stateexp}:
\begin{gather}
	|\phi, \theta, \lambda_a, \lambda_b, |\bold{p}|\rangle 
	= \sum_{J,M} C_J\mathscr{D}^J_{M, \lambda_a-\lambda_b}(R(\phi, \theta, 0)) |J, M, \lambda_a, \lambda_b, |\bold{p}|\rangle.
\label{eq:rotstateexp}	 
\end{gather}
The inverse expansion can be obtained using the properties of the $\mathscr{D}$-matrices:
\begin{gather}
	|J, M, \lambda_a, \lambda_b, |\bold{p}|\rangle = C_J\int_0^{2\pi}d\phi \int_0^{\pi}d\theta \sin \theta \mathscr{D}^{J*}_{M, \lambda_a-\lambda_b}(R(\phi, \theta, 0))|\phi, \theta, \lambda_a, \lambda_b, |\bold{p}|\rangle.
\label{eq:invrotstateexp}	 
\end{gather}
This can be used to compute perturbatively the amplitude for the production of a vector boson from a state with a particular $J$:
\begin{gather}
	\langle 0 | J^{\mu}(0) |J, M, \lambda_a, \lambda_b, |\bold{p}|\rangle = C_J\int_0^{2\pi}d\phi \int_0^{\pi}d\theta \sin \theta \mathscr{D}^{J*}_{M, \lambda_a-\lambda_b}(R(\phi, \theta, 0))\langle 0 | J^{\mu}(0)|\phi, \theta, \lambda_a, \lambda_b, |\bold{p}|\rangle,
\label{eq:rotstateamp1}	 
\end{gather}
or, for a full process with the final state $| f \rangle$,
\begin{gather}
	\langle f |J, M, \lambda_a, \lambda_b, |\bold{p}|\rangle = C_J\int_0^{2\pi}d\phi \int_0^{\pi}d\theta \sin \theta \mathscr{D}^{J*}_{M, \lambda_a-\lambda_b}(R(\phi, \theta, 0))\langle f |\phi, \theta, \lambda_a, \lambda_b, |\bold{p}|\rangle.
\label{eq:rotstateamp2}	 
\end{gather}
The Feynman rules for the momentum eigenstates on the r.h.s. of Eqs.~\ref{eq:rotstateamp1},~\ref{eq:rotstateamp2} are well known.
In the following we shall compute a few simple processes to show the propagation of $J$ values.

\section{Example processes}
\label{sec:ex_proc}
 
\begin{figure}[!ht]
  \centering
	\begin{subfigure}[t!]{0.4\linewidth}
	  \parbox[][4cm][c]{\linewidth}{
      \centering
	  \includegraphics[scale=0.5]{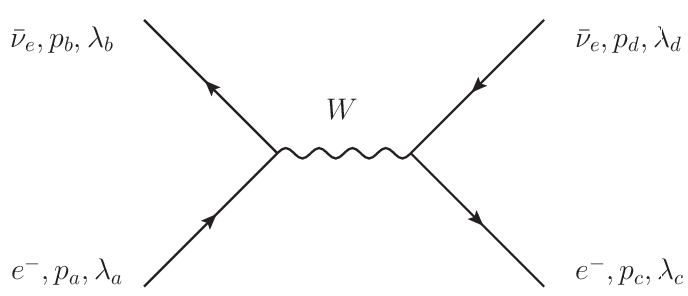}
	}
	  \subcaption{$e^-_L \bar{\nu}_{eL} \rightarrow W \rightarrow e^-_L \bar{\nu}_{eL}$.}
  \end{subfigure}
  \begin{subfigure}[t!]{0.4\linewidth}
	  \parbox[][4cm][c]{\linewidth}{
      \centering
	  \includegraphics[scale=0.5]{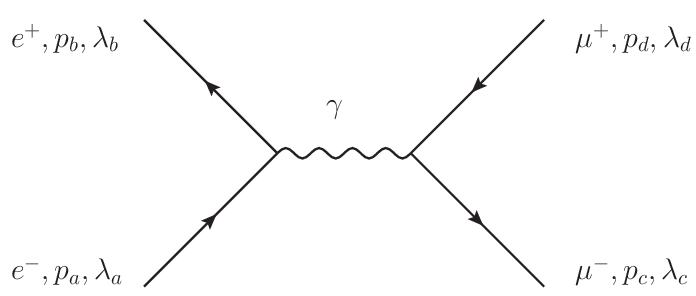}
	  }
	  \subcaption{$e^-_L e^+_L \rightarrow \gamma \rightarrow \mu^-_L \mu^+_L$.}
  \end{subfigure}
	\caption{Example 2 to 2 processes at tree level with the production and decay of a vector boson, where the fermion polarizations
	are defined by their helicities given in the main text.}
\label{fig:2to2}
\end{figure}
We consider 2 to 2 tree level fermion scattering processes in $s$-channel mediated by a vector boson.  
The two processes in Fig.~\ref{fig:2to2} will be treated in sequence.
In these calculations, we would like to see explicitly the contribution
from the $J=0$ and $J=1$ eigenstates, in accord with the generic result obtained in section~\ref{sec:gen_disc}. 
Therefore the initial and final state fermions are taken to have definite helicities
such that the angular structure of the processes are easily revealed. Specifically, we require that the two fermions in the initial 
state have the opposite spin components along the direction of the incoming beams. This is necessary to ensure that the $J=0$ partial
wave is non-zero in the expansion of the momentum eigenstates in Eq.~\ref{eq:stateexp}. In contrast, if the fermions have the same spin along the
beam, then the total spin of the system is $S=1$, which cannot be canceled by the orbital angular momentum to give $J=0$ because the orbital 
angular momentum is perpendicular to the beam direction. In this case, the partial wave expansion only contains terms with $J\geq 1$.

 \subsection{$e^-_L(p_a) \bar{\nu}_{eL}(p_b) \rightarrow W \rightarrow e^-_L(p_c) \bar{\nu}_{eL}(p_d)$}

Here the subscript ``L'' for all the four fermions refers to the helicity configuration $\lambda_a=-\lambda_b=\lambda_c=-\lambda_d=-1$,  
rather than  definite chiralities. In other words, we use ``L'' and ``R'' to describe the relation between 
the direction of a particle's physical spin and the direction in which it moves. This choice of polarizations is not 
contradicting to the fact that the anti-neutrino can only be right-handed in chirality, as long as it is a massive particle.
Recall that when obtaining $M=\lambda_a-\lambda_b$ we give an extra minus sign to $\lambda_b$ for the
anti-neutrino, so that in this case $M=0$ instead of $-2$.
In the massless limit, the helicity and chirality eigenstates coincide, but this amplitude vanishes by helicity conservation. However,
we shall keep the fermion masses so that the $J=0$ mode is allowed, as will be seen below. 
Recall that we are examining the propagation of each $J$-mode in principle 
rather than evaluating the practical importance of the contribution from each $J$ (which will be discussed in section~\ref{sec:pheno}). 

The initial momenta are
\begin{gather}
	p_a = (E_a, 0, 0, p), \quad E_a = \sqrt{m_a^2+p^2}, \\
	p_b = (E_b, 0, 0, -p),\quad E_b = \sqrt{m_b^2+p^2},
\label{eq:inimom_enu}	 
\end{gather}
where $m_a\equiv m_e$, and $m_b\equiv m_{\nu_e}$.
The tree level vertex for the subprocess $e^-_L(p_a) \bar{\nu}_{Le}(p_b) \rightarrow W$ is (excluding the W propagator and an overall factor
$-i\sqrt{2\pi\alpha}/\sin \theta_W$ at the vertex)
\begin{gather}
	V^{\mu}(e^-_L(p_a) \bar{\nu}_{eL}(p_b) \rightarrow W)=\langle 0 | J^{\mu}(0) | p_a,\lambda_a; p_b, \lambda_b \rangle_{\text{tree}}= \bar{v}^{\lambda_b}(p_b)\gamma^{\mu}\frac{1-\gamma^5}{2}u^{\lambda_a}(p_a), 
\label{eq:enuVertini}	 
\end{gather}
where $J^{\mu}(0)=\overline{\Psi^{\nu_e}}\gamma^{\mu}\frac{1-\gamma_5}{2}\Psi^{e}$, 
and the particle species are omitted in the states and spinors. Here we follow
the method in Ref.~\cite{Peskin:1995ev} to compute the amplitude for polarized states and employ the Weyl representation 
for the Dirac $\gamma$-matrices: 
\begin{gather}
\gamma^0=
\begin{pmatrix}
0 & 1   \\
1 & 0   
\end{pmatrix}
,\quad
\gamma^i=
\begin{pmatrix}
0 & \sigma_i   \\
-\sigma_i & 0   
\end{pmatrix}
,\quad
\gamma^5=
\begin{pmatrix}
-1 & 0   \\
0 & 1   
\end{pmatrix},
\label{eq:gammamats}	 
\end{gather}
in which each element is a 2 by 2 matrix and $\sigma_i$ are the Pauli matrices. 
The corresponding spinors for the spin-up electron and spin-down anti-neutrino (both with respect to the 3-direction) are
\begin{gather}
	[u^{\lambda_a}(p_a)]^T = (0, \sqrt{E_a+p}, 0, \sqrt{E_a-p}), \\
	[v^{\lambda_b}(p_b)]^T = (0, \sqrt{E_b-p}, 0, \sqrt{E_b+p}).
\label{eq:spinors_enu}	 
\end{gather}
Inserting these expressions into Eq.~\ref{eq:enuVertini} gives
\begin{gather}
	V^{\mu}(e^-_L(p_a) \bar{\nu}_{eL}(p_b) \rightarrow W)= \sqrt{(E_b-p)(E_a+p)}(1,0,0,1).
\label{eq:enuVertini2}	 
\end{gather}

The vertex for $W \rightarrow e^-_L(p_c) \bar{\nu}_{eL}(p_d)$ is
\begin{gather}
	V^{\mu}(W \rightarrow e^-_L(p_c) \bar{\nu}_{eL}(p_d))= \bar{u}^{\lambda_c}(p_c)\gamma^{\mu}\frac{1-\gamma^5}{2}v^{\lambda_d}(p_d) 
	=\left [\bar{v}^{\lambda_d}(p_d)\gamma^{\mu}\frac{1-\gamma^5}{2}u^{\lambda_c}(p_c)\right ]^*.
\end{gather}
The term in the square brackets is of the same form as in Eq.~\ref{eq:enuVertini}, except the momenta $\bold{p}_a =-\bold{p}_b$ are rotated
from the 3-direction to $\bold{p}_c =-\bold{p}_d$. 
This term must be a 4-vector that transforms the same way as $p_a$ and $p_b$ under the rotation $\Lambda(R)$ that rotates them to $p_c$ and $p_d$. 
This simply follows from the transformation of the vertex function
\begin{align}
	&\langle 0 | J^{\mu}(0) | p_c,\lambda_a; p_d, \lambda_b \rangle \nonumber \\
	&=\langle 0 |U(\Lambda(R))U^{-1}(\Lambda(R)) J^{\mu}(0) U(\Lambda(R))U^{-1}(\Lambda(R))| p_c,\lambda_a; p_d, \lambda_b \rangle, \nonumber \\
	&=\Lambda(R)^{\mu}_{\nu}\langle 0 | J^{\nu}(0) | \Lambda^{-1}(R)p_c,\lambda_a; \Lambda^{-1}(R)p_d, \lambda_b \rangle \nonumber \\
	&=\Lambda(R)^{\mu}_{\nu}\langle 0 | J^{\nu}(0) | p_a,\lambda_a; p_b, \lambda_b \rangle,
	\label{eq:tensorform}	 
\end{align}
which gives\footnote{The tensor structure on the r.h.s. of Eq.~\ref{eq:piVert} 
can be deduced in a similar way.} immediately at tree level
\begin{gather}
	V^{\mu}(W \rightarrow e^-_L(p_c) \bar{\nu}_{eL}(p_d))=\left [\Lambda(R)^{\mu}_{\nu}V^{\mu}(e^-_L(p_a) \bar{\nu}_{eL}(p_b) \rightarrow W) \right ]^*= 
	\sqrt{(E_b-p)(E_a+p)}(1,\bold{p}_c/p),
\label{eq:enuVertfin}	 
\end{gather}
where $E_a=E_c$ and $E_b=E_d$. 

Now restore the couplings (and other overall factors) at the vertices 
and contract with the W propagator in unitary gauge\footnote{We would like to show that the propagation of zero angular momentum is possible
without resorting to the unphysical Goldstone boson.}
\begin{gather}
	\frac{-i}{q^2-M_W^2}(g_{\mu \nu}-\frac{q_{\mu}q_{\nu}}{M_W^2}) \\
	(q\equiv p_a+p_b) \nonumber
\end{gather}
to get the amplitude for the full process
\begin{gather}
	\mathcal{M}(e^-_L(p_a) \bar{\nu}_{eL}(p_b) \rightarrow W \rightarrow e^-_L(p_c) \bar{\nu}_{eL}(p_d))= 
	\frac{2\pi \alpha}{\sin^2 \theta_W}\frac{(E_b-p)(E_a+p)}{(E_a+E_b)^2-M_W^2}\left[1-\frac{(E_a+E_b)^2}{M_W^2}-\cos \theta \right],
	\label{eq:enuamp}	 
\end{gather}
where $M_W$ is the mass of the W boson. 
The dependence of the result on the external momenta is only through the angle $\theta$ between the initial and final fermion beams.

It is straightforward to compute the amplitude with an angular momentum eigenstate using Eq.~\ref{eq:rotstateamp2}, in the form
\begin{align}
	&\langle \phi, \theta, \lambda_c=\lambda_a, \lambda_d=\lambda_b, p |J, M, \lambda_a, \lambda_b, p\rangle \nonumber \\
	&= C_J\int_0^{2\pi}d\phi^{\prime} \int_0^{\pi}d\theta^{\prime} \sin \theta^{\prime} \mathscr{D}^{J*}_{M, \lambda_a-\lambda_b}(R(\phi^{\prime}, \theta^{\prime}, 0))\langle \phi, \theta, \lambda_c=\lambda_a, \lambda_d=\lambda_b, p |\phi^{\prime}, \theta^{\prime}, \lambda_a, \lambda_b, p\rangle.
\label{eq:rotstateamp3}	 
\end{align}
The matrix element on the r.h.s. of Eq~\ref{eq:rotstateamp3} is computed with the initial and final states
\begin{gather}
	| i \rangle \equiv| p_a^{\prime},\lambda_a; p_b^{\prime}, \lambda_b \rangle = |\phi^{\prime}, \theta^{\prime}, \lambda_a, \lambda_b, |\bold{p}_a^{\prime}|=p \rangle, \\
    | f \rangle \equiv| p_c,\lambda_c; p_d, \lambda_d \rangle = |\phi, \theta, \lambda_c=\lambda_a, \lambda_d=\lambda_b, |\bold{p}_c|=p \rangle, \\
	\langle f | i \rangle_{\text{tree}} = \langle p_c,\lambda_c; p_d, \lambda_d| p_a^{\prime},\lambda_a; p_b^{\prime}, \lambda_b \rangle_{\text{tree}} = N_{ab}\left[1-\frac{(E_a+E_b)^2}{M_W^2}-\cos \theta_{ac}\right],
	\label{eq:enuamp1}	 
\end{gather}
where on the r.h.s. of Eq.~\ref{eq:enuamp1} (and below), we suppress the factor $i(2\pi)^4$ and the delta function that implements 
the overall momentum conservation. $N_{ab}$ represents the  coefficient in Eq.~\ref{eq:enuamp} multiplying the 
square bracket, and $\theta_{ac}$ is the angle between $\bold{p}_a^{\prime}$ and $\bold{p}_c$:
\begin{gather}
	\cos \theta_{ac} = \sin \theta  \sin \theta^{\prime} \cos (\phi- \phi^{\prime})+\cos \theta \cos \theta^{\prime}.
	\label{eq:costhetaac}	 
\end{gather}
There is no contradiction between Eq.~\ref{eq:enuamp1} and Eq.~\ref{eq:momstatesnorm}, because Eq.~\ref{eq:enuamp1} 
is a product of ``in'' and ``out'' states that makes a transition amplitude in the presence of interactions, 
while Eq.~\ref{eq:momstatesnorm} is a normalization condition when both
states are ``in'', ``out'', or free states. We do not label these states with 
``in'', ``out'', or ``free'' since they can be distinguished by context. 

Inserting the explicit expressions for the $\mathscr{D}$-matrices~\cite{Jacob:1959at,Leader:2011vwq}, the angular integration of 
Eq.~\ref{eq:rotstateamp3} can be trivially performed to give
\begin{align}
	\langle \phi, \theta, \lambda_c=\lambda_a, \lambda_d=\lambda_b, p |J, M, \lambda_a, \lambda_b, p\rangle_{\text{tree}}= 
	N_{ab}\times \left\{
		\begin{aligned}
			&-\sqrt{\frac{4\pi}{3}}\cos \theta,\,&\text{for} \, J=1, M=0,\\
			&\sqrt{4\pi}\left[1-\frac{(E_a+E_b)^2}{M_W^2}\right],\,&\text{for} \, J=0, M=0.
		\end{aligned}
		\right.
\label{eq:enuamp2}	 
\end{align}
Hence, both $J=0$ and $J=1$ partial waves contribute to the amplitude Eq.~\ref{eq:enuamp}. \footnote{The smallness of the neutrino mass $m_b$
makes the factor $E_b-p$ close to zero in the expression for the amplitude (which does not affect our arguments of course). 
If one chooses to compute alternatively the amplitude for the
same process but with all external fermions right-handed, a factor $E_a-p$ will appear instead of $E_b-p$, and the non-zero
contribution is still from $J=0$ and $J=1$.}
Indeed, plugging the result of Eq.~\ref{eq:enuamp2} into the 
expansion Eq.~\ref{eq:matexp2} reproduces Eq.~\ref{eq:enuamp} immediately. 
Note the $J\geq 2$ contribution vanishes. This can be seen from the form of the terms on the r.h.s. of Eq.~\ref{eq:enuamp1}.
Those terms proportional to $1$ and $\cos \theta^{\prime}$
can be expressed by the matrix elements $\mathscr{D}^{J=0}_{00}$ and $\mathscr{D}^{J=1}_{00}$, which are orthogonal to the $\mathscr{D}^{J*}_{00}$ with 
$J\geq 2$ in the integration in Eq.~\ref{eq:rotstateamp3}.  
The other terms are proportional to eigher $\cos \phi^{\prime}$ or $\sin \phi^{\prime}$,
which vanish after the integral over $\phi^{\prime}$. This is consistent with our general argument that states with $J\geq 2$
transform differently from the transformation of the current under rotations. Note that the argument was made for the vertex in Fig.~\ref{fig:Vert} (a),
with all order contributions. However, the vanishing of the vertex must be order by order since the perturbation series is zero for arbitrary
values of the coupling constants involved in the interactions.

 \subsection{$e^-_L(p_a) e^+_L(p_b) \rightarrow \gamma \rightarrow \mu^-_L(p_c) \mu^+_L(p_d)$}
 
 A calculation similar to the one carried out above gives
\begin{gather}
	\mathcal{M}(e^-_L(p_a) e^+_L(p_b) \rightarrow \gamma \rightarrow \mu^-_L(p_c) \mu^+_L(p_d))= 
	-\frac{4\pi \alpha m_a m_c}{E_a^2 }\cos \theta,
	\label{eq:eeamp}	 
\end{gather}
where $m_a=m_e$ and $m_c=m_{\mu}$. In place of Eq.~\ref{eq:enuamp2} we have
\begin{align}
	\langle \phi, \theta, \lambda_c=\lambda_a, \lambda_d=\lambda_b, p |J, M, \lambda_a, \lambda_b, p\rangle_{\text{tree}}= 
	N_{ab}\times \left\{
		\begin{aligned}
			&-\sqrt{\frac{4\pi}{3}}\cos \theta,\,&\text{for} \, J=1, M=0,\\
			&\qquad 0,\,&\text{for} \, J=0, M=0,
		\end{aligned}
		\right.
\label{eq:eeamp2}	 
\end{align}
where now $N_{ab}$ stands for the factor multiplying $-\cos \theta$ in Eq.~\ref{eq:eeamp}. This result shows that only the $J=1$ mode
contributes to the angular distribution of the final state, which by itself reproduces Eq.~\ref{eq:eeamp}. The amplitudes with $J\geq 2$
are zero by the similar arguments given above. The new feature compared to the $e^-$-$\bar{\nu}_{e}$ scattering is the vanishing of the $J=0$ result,
which is not a coincidence and has a deeper origin. 

To see this we note that a positronium state with definite spin $S$ and orbital angular momentum $L$ is
an eigenstate of the charge conjugation operator $C$, with the eigenvalue $\eta_C=(-1)^{L+S}$~\cite{Perkins:2000xb,Fayyazuddin:1994wh}. 
This property can be 
carried over to analyzing our process with an unbounded electron-positron pair. 
The state $|J, M, \lambda_a, \lambda_b, p\rangle$ with $J=M=0$ has a definite $J$ 
rather than $L$ or $S$. However, the rule of adding two angular momenta requires that $J=0$ can only be obtained 
by combining states with definite $L$ and $S$ that satisfy $L=S$.
For all these states, $(-1)^{L+S}=1$, 
and $|J=0, M=0\rangle$ is therefore a state that is even under charge conjugation:
\begin{gather}
    C |J=0, M=0\rangle = |J=0, M=0\rangle,
\end{gather}
where the quantum numbers other than $J$ and $M$ are suppressed in the state. Since the electromagnetic current is C-odd, we have
\begin{align}
	&\langle 0 | J^{\mu}(0) |J=0, M=0\rangle \nonumber\\
	&=\langle 0 |C^{-1}C J^{\mu}(0)C^{-1}C |J=0, M=0\rangle \nonumber\\
	&=-\langle 0 | J^{\mu}(0) |J=0, M=0\rangle \nonumber\\
	&= 0,
\label{eq:Ctransfmat}	 
\end{align}
which merely reflects the fact that $J^{\mu}$ and $|J=0, M=0\rangle$ 
belong to different representations of the charge conjugation operation (one C-even and the other C-odd).
So the $J=0$ mode does vanish.\footnote{The result here and in section~\ref{sec:gen_disc} are obtained for the vertex functions. 
However, as we are focusing on $s$-channel graphs with a vector boson, these vertices are parts of the amplitudes.
Furthermore, even though the contribution from $J^{\mu}$, with $\mu=1,2,$ or $3$ is also zero by rotational invariance of the states 
, this does not ensure they are zero outside the CM frame of the electron-positron pair (where the rotational invariance of the initial state is lost. 
See Eq.~\ref{eq:piVert} for a similar instance, where the spatial components of the matrix element is non-zero outside the pion rest frame.), 
unless the contribution from $J^0$ is also zero.}
Apparently, this argument cannot be applied to the case of $e^-$-$\bar{\nu}_e$ scattering or charged pion decay,
because the initial state of either process has a net electric charge that is reversed by C, and is therefore not an eigenstate of C.
The initial state must then have a non-zero projection onto the C-odd space of the states, 
such that the amplitude with the current is not restricted to zero by charge conjugation.

\section{Phenomenology}
\label{sec:pheno}

Given a non-zero $J=0$ mode carried by a massive gauge boson in processes from previous sections, 
the immediate question about the role of this mode arises when 
determining the spin of the vector boson from experiments. For instance, the spin of the W boson was determined by measurement of the 
angular distribution of the electrons or positrons from the decay $W\rightarrow e\nu_e$ 
in the experiment carried out by the UA1 collaboration~\cite{UA1:1985bfo,UA1:1985our,UA1:1988rck}, in which W bosons were produced via Drell-Yan like
processes in proton-antiproton collisions. The measured unpolarized\footnote{The filtering of various polarization combinations is only done through the 
nature of the $V-A$ interactions.} distribution is consistent with the shape
$(1+\cos \theta)^2$, where $\theta$ is the angle between the directions of the electron (positron) and the proton (antiproton) in the rest
frame of the W boson. The shape of the distribution can be traced back to the Wigner 
$\mathscr{D}$-matrix $\mathscr{D}^{1*}_{1,1}(R(\phi, \theta, 0))=e^{-i\phi}(1+\cos \theta)/2$. It shows that the W boson is produced in a state
with $J=M=1$.  

The scalar mode\footnote{The scalar and vector modes in this section refer to the behaviors of the states under spatial rotations.} 
of the W boson contributes an isotropic term to the distribution of the decay product (see Eq.~\ref{eq:enuamp2}). 
This would in principle distort the distribution deduced from the vector mode of the W. However,  
the $V-A$ form only allows left-handed fermions and right-handed antifermions
to participate in the interaction in the massless limit, in which case the propagation of the scalar mode is forbidden.  
In fact, it is straightforward to show that a fermion (antifermion) with
a right-handed (left-handed) spin would introduce a factor $\sqrt{E-|\bold{p}|}=m/\sqrt{E+|\bold{p}|}$ to the scattering amplitude, 
where $E$, $\bold{p}$, and $m$ are the energy, 3-momentum, and mass of the fermion (antifermion) (See Eqs.~\ref{eq:enuVertini2} 
and ~\ref{eq:enuVertfin} for instance). If $m\ll E$, this gives a suppression factor 
$m/(2E)$ (known as helicity suppression) relative to the amplitude 
that is not suppressed (e.g., the one proportional to $\mathscr{D}^1_{1,1}(R(\phi, \theta, 0))$ above). 
Here $E$ is of the order of the parton level CM energy for the scattering. 

For an estimate with the UA1 experiment we could take $E\sim 100$ GeV, and consider the scalar mode that
propagates via the process $d_R \bar{u}_{R} \rightarrow W \rightarrow \mu^-_R \bar{\nu}_{\mu R}$ (The process with left handed spins is
much smaller since the final state suppression factor is from the mass of the neutrino rather than from that of the muon. Also we consider
the muon instead of the electron in the final state since it has a larger mass). The suppression
factor at the cross section level is $\sim \mathcal{O}((\frac{m_d m_{\mu}}{4E^2})^2) \sim 10^{-16}$. This is far too small to be observed. 
The statistical uncertainty of the measured angular distribution in the UA1 experiment is as large as 50\%. For total cross section and
forward-backward asymmetry observables, the accuracy improves to $\sim 10\%$. More recent measurements at the 
Fermilab Tevatron~\cite{D0:2000ydn,CDF:2003tlc} and the LHC~\cite{ATLAS:2011qdp,CMS:2020cph} reached a percent or even per-mille level accuracy, 
but are still far from being sensitive to the mass-suppressed modes.

To further reduce the mass-suppression effect on the scalar mode, we might consider the $s$-channel single top production, so that the 
final state suppression factor comes from the top quark mass. However, not much could be done about the initial state. Even if we consider
the ``sea-sea'' contribution to the production of the W boson and take the charm quark from an incoming beam, we get a factor 
$\sim \mathcal{O}((\frac{m_c m_t}{4E^2})^2) \sim 10^{-6}$, where $E\sim 300$ GeV for the Tevatron experiment. This will further be suppressed by the 
sea quark distributions. Also the identification of the top final state is much more difficult than the leptonic decay product of the W~\cite{D0:2013tnv,CDF:2014uma,ATLAS:2015jmq,CMS:2016xoq}.

At this point it appears hopeless to probe the $J=0$ state of a massive vector boson produced via 
Drell-Yan (like) processes.\footnote{Note that not only the scalar mode, but also
the one with $J=1$, $M=0$ (see Eq.~\ref{eq:enuamp2}) is suppressed. However, this latter mode also propagates 
via a final state with no suppression factor, e.g.,  $d_R \bar{u}_{R} \rightarrow W \rightarrow \mu^-_L \bar{\nu}_{\mu R}$. 
(The suppression factor is only from the initial state.)
A similar estimate gives a somewhat larger result for this process compared with the scalar mode. 
But the improvement is still very limited compared to the uncertainties of the measurements.} 
It is out of the scope of this study to perform a scan of processes, but here we leave the possibility open 
that this scalar state could manifest in some other production modes of the vector bosons, in addition to the decay processes of
e.g., the charged pions.

\section{Summary and discussion}

In this study, we analyze the propagation of angular momentum via a vector boson in several processes by expanding the initial states
into a superposition of angular momentum eigenstates of the system. Through a general discussion 
based on symmetry principles and two practical calculations, we are able to show that both the singlet ($J=0$) and 
triplet ($J=1$) representations of the rotation group can be transferred to the final state via an intermediate vector boson, 
unless constraints from further symmetries are present to forbid it. However, we have shown that the scalar state of a massive vector boson
is produced with an extremely small rate that it has no impact on existing experiments at hadron colliders. The decays of several spin-zero
particles, such as the charged pions, are the only obvious phenomena we have found related with the scalar state of gauge bosons.

The crucial point to understand the $J=0$ case is that an initial state that is invariant under spatial rotations is not a singlet under a 
generic Lorentz transformation. The $J=0$ state can be changed by boosts in any direction and gives a non-zero matrix element for producing
a Lorentz vector current to which a massive gauge boson couples. Hence there is no 
Lorentz violation in the production of an off-shell gauge boson from such
a state via interactions that are Lorentz invariant. The spin of a vector boson as a virtual particle has to be measured via its decay product and is
not necessarily equal to its spin as a free field.

On the other hand, an on-shell vector boson should only be in a state with spin 1, and this should also be reflected at the amplitude level. 
We have shown that a state with $J=M=0$ can propagate though the $\mu =0$ component of a vector current, as appears in Eq.~\ref{eq:matexp}. However,
this fact alone does not ensure the full amplitude with a $J=M=0$ initial state is none zero, 
since the initial vertex function will be contracted with the rest part of the amplitude via the propagator of a gauge boson.
For instance, we may think of the decay amplitude for $\pi^-\rightarrow l^- \bar{\nu}_l$ at leading order.
After accounting for Eq.~\ref{eq:piVert}, the amplitude takes the form 
\begin{gather}
	\mathcal{M}(\pi^-(k)\rightarrow l^-(k_1) \bar{\nu}_l(k_2)) \sim k^{\mu}\frac{-g_{\mu \nu}+\frac{k_{\mu}k_{\nu}}{M_W^2}}{k^2-M_W^2+i\epsilon} 
        \bar{u}(k_1)\gamma^{\nu}\frac{1-\gamma^5}{2}v(k_2)
\label{eq:pidecLO} 
\end{gather}
where the tensor structure of the propagator in unitary gauge is obtained on the W mass shell by 
summing over three physical polarization vectors of the W boson
\begin{gather}
	\sum_{\epsilon_{\mu}k^{\mu}=0}\epsilon_{\mu}\epsilon^*_{\nu} = -g_{\mu \nu}+\frac{k_{\mu}k_{\nu}}{M_W^2}.
\label{eq:polvecsum} 
\end{gather}
These polarization vectors are purely spatial in the W rest frame and make an $SO(3)$ triplet under rotations. 
Therefore the contraction of the propagator with $k^{\mu}$ (whose spatial components are all zero) in Eq.~\ref{eq:pidecLO} gives
\begin{gather}
	k^{\mu}\frac{-g_{\mu \nu}+\frac{k_{\mu}k_{\nu}}{M_W^2}}{k^2-M_W^2+i\epsilon} = \frac{k_{\nu}}{M_W^2}\frac{k^2-M_W^2}{k^2-M_W^2+i\epsilon}
     = \left\{
		\begin{aligned}
			&\quad 0,\,& \, k^2=M_W^2,\\
			&k_{\nu}/M_W^2,\,& \, k^2 \ne M_W^2.
		\end{aligned}
		\right.
\label{eq:pidecLOctrct} 
\end{gather}
Of course, experimentally the pion mass $m_{\pi^-}$ is much smaller than $M_W$. But as $M_W$ is a free parameter of the SM, we allow $k^2$ to get
on the W mass shell (i.e., $m_{\pi^-}=M_W$) for the purpose of checking the consistency of the calculation.
The infinitesimal imaginary part of the denominator in Eq.~\ref{eq:pidecLOctrct} then ensures that 
the contraction is zero. This structure of contraction persists in loop corrections of the decay amplitude to all orders. 
Therefore an on-shell W boson indeed forbids the propagation of a $J=0$ state. In the off-shell case, however, the state of the W boson
cannot be described by the polarization vectors in Eq.~\ref{eq:polvecsum} that furnish a $\mathscr{D}^1$ representation of the rotation group,
and hence allows for a non-zero decay rate. This discontinuous transition of the decay amplitude to zero at the W mass shell shows clearly the 
consistency of SM dynamics with angular momentum conservation\footnote{Another example would be from the $J=0$ result in 
Eq~\ref{eq:enuamp2}, when $E_a+E_b$ is getting on shell: $E_a+E_b \to M_W$. Because of the propagator denominator from $N_{ab}$, the transition
of the $J=0$ mode to zero occurs in exactly the same way as with the pion decay amplitude here.}.

\section{Acknowledgement}

This work was supported by National Natural Science Foundation of China (12105068) and Hangzhou Normal University Start-up Funds.

\section*{Data availability statement}
There is no data associated with this study.

\bibliographystyle{utphysmcite}
\bibliography{PiDec}

\ifx\mcitethebibliography\mciteundefinedmacro
\PackageError{unsrtM.bst}{mciteplus.sty has not been loaded}
{This bibstyle requires the use of the mciteplus package.}\fi
\begin{mcitethebibliography}{10}

\bibitem{Peskin:1995ev}
M.~E. Peskin and D.~V. Schroeder, {\em {An Introduction to quantum field
  theory}}.
\newblock Addison-Wesley, Reading, USA\relax
\mciteBstWouldAddEndPunctfalse
\mciteSetBstMidEndSepPunct{\mcitedefaultmidpunct}
{}{\mcitedefaultseppunct}\relax
\EndOfBibitem
\bibitem{Weinberg:1995mt}
S.~Weinberg, \href{http://dx.doi.org/10.1017/CBO9781139644167}{{\em {The
  Quantum theory of fields. Vol. 1: Foundations}}}.
\newblock Cambridge University Press\relax
\mciteBstWouldAddEndPunctfalse
\mciteSetBstMidEndSepPunct{\mcitedefaultmidpunct}
{}{\mcitedefaultseppunct}\relax
\EndOfBibitem
\bibitem{Sterman:1993hfp}
G.~F. Sterman, {\em {An Introduction to quantum field theory}}.
\newblock Cambridge University Press\relax
\mciteBstWouldAddEndPunctfalse
\mciteSetBstMidEndSepPunct{\mcitedefaultmidpunct}
{}{\mcitedefaultseppunct}\relax
\EndOfBibitem
\bibitem{ParticleDataGroup:2022pth}
{Particle Data Group}, R.~L. Workman {\em et~al.}, ``{Review of Particle
  Physics},'' \href{http://dx.doi.org/10.1093/ptep/ptac097}{{\em PTEP}
  {\bfseries 2022} (2022) 083C01}\relax
\mciteBstWouldAddEndPunctfalse
\mciteSetBstMidEndSepPunct{\mcitedefaultmidpunct}
{}{\mcitedefaultseppunct}\relax
\EndOfBibitem
\bibitem{Pocanic:2014jka}
D.~Pocanic, E.~Frlez, and A.~van~der Schaaf, ``{Experimental study of rare
  charged pion decays},''
  \href{http://dx.doi.org/10.1088/0954-3899/41/11/114002}{{\em J. Phys. G}
  {\bfseries 41} (2014) 114002},
  \href{http://arxiv.org/abs/1407.2865}{{\ttfamily arXiv:1407.2865
  [hep-ex]}}\relax
\mciteBstWouldAddEndPunctfalse
\mciteSetBstMidEndSepPunct{\mcitedefaultmidpunct}
{}{\mcitedefaultseppunct}\relax
\EndOfBibitem
\bibitem{Abashian:1957zz}
A.~Abashian {\em et~al.}, ``{Angular Distributions of Positrons from pi+-mu+-e+
  Decays Observed in a Liquid Hydrogen Bubble Chamber},''
  \href{http://dx.doi.org/10.1103/PhysRev.105.1927}{{\em Phys. Rev.} {\bfseries
  105} (1957) 1927--1928}\relax
\mciteBstWouldAddEndPunctfalse
\mciteSetBstMidEndSepPunct{\mcitedefaultmidpunct}
{}{\mcitedefaultseppunct}\relax
\EndOfBibitem
\bibitem{PhysRevLett.14.745}
S.~Taylor, E.~L. Koller, T.~Huetter, P.~Stamer, and J.~Grauman, ``Search for
  anomalous ${\ensuremath{\pi}}^{+}$ decay among ${\ensuremath{\tau}}^{+}$
  decay secondaries,'' \href{http://dx.doi.org/10.1103/PhysRevLett.14.745}{{\em
  Phys. Rev. Lett.} {\bfseries 14} (May, 1965) 745--746}.
  \url{https://link.aps.org/doi/10.1103/PhysRevLett.14.745}\relax
\mciteBstWouldAddEndPunctfalse
\mciteSetBstMidEndSepPunct{\mcitedefaultmidpunct}
{}{\mcitedefaultseppunct}\relax
\EndOfBibitem
\bibitem{Frota-Pessoa:1969jeg}
E.~Frota-Pessoa, ``{Isotropy in pi-minus mu decays},''
  \href{http://dx.doi.org/10.1103/PhysRev.177.2368}{{\em Phys. Rev.} {\bfseries
  177} (1969) 2368--2370}\relax
\mciteBstWouldAddEndPunctfalse
\mciteSetBstMidEndSepPunct{\mcitedefaultmidpunct}
{}{\mcitedefaultseppunct}\relax
\EndOfBibitem
\bibitem{Hulubei:1963zza}
H.~Hulubei, J.~S. Auslander, E.~M. Friedlander, and S.~Titeica, ``{Angular
  Distribution of Muons in pi-mu Decay at Rest},''
  \href{http://dx.doi.org/10.1103/PhysRev.131.2841}{{\em Phys. Rev.} {\bfseries
  129} (1963) 2789--2801}. [Erratum: Phys.Rev. 131, 2841 (1963)]\relax
\mciteBstWouldAddEndPunctfalse
\mciteSetBstMidEndSepPunct{\mcitedefaultmidpunct}
{}{\mcitedefaultseppunct}\relax
\EndOfBibitem
\bibitem{MINOS:2008fnv}
{MINOS}, P.~Adamson {\em et~al.}, ``{Testing Lorentz Invariance and CPT
  Conservation with NuMI Neutrinos in the MINOS Near Detector},''
  \href{http://dx.doi.org/10.1103/PhysRevLett.101.151601}{{\em Phys. Rev.
  Lett.} {\bfseries 101} (2008) 151601},
  \href{http://arxiv.org/abs/0806.4945}{{\ttfamily arXiv:0806.4945
  [hep-ex]}}\relax
\mciteBstWouldAddEndPunctfalse
\mciteSetBstMidEndSepPunct{\mcitedefaultmidpunct}
{}{\mcitedefaultseppunct}\relax
\EndOfBibitem
\bibitem{MINOS:2012ozn}
{MINOS}, P.~Adamson {\em et~al.}, ``{Search for Lorentz invariance and CPT
  violation with muon antineutrinos in the MINOS Near Detector},''
  \href{http://dx.doi.org/10.1103/PhysRevD.85.031101}{{\em Phys. Rev. D}
  {\bfseries 85} (2012) 031101},
  \href{http://arxiv.org/abs/1201.2631}{{\ttfamily arXiv:1201.2631
  [hep-ex]}}\relax
\mciteBstWouldAddEndPunctfalse
\mciteSetBstMidEndSepPunct{\mcitedefaultmidpunct}
{}{\mcitedefaultseppunct}\relax
\EndOfBibitem
\bibitem{Altschul:2013yja}
B.~Altschul, ``{Contributions to Pion Decay from Lorentz Violation in the Weak
  Sector},'' \href{http://dx.doi.org/10.1103/PhysRevD.88.076015}{{\em Phys.
  Rev. D} {\bfseries 88} (2013) 076015},
  \href{http://arxiv.org/abs/1308.2602}{{\ttfamily arXiv:1308.2602
  [hep-ph]}}\relax
\mciteBstWouldAddEndPunctfalse
\mciteSetBstMidEndSepPunct{\mcitedefaultmidpunct}
{}{\mcitedefaultseppunct}\relax
\EndOfBibitem
\bibitem{Altschul:2013ykb}
B.~Altschul, \href{http://dx.doi.org/10.1103/PhysRevD.87.096004}{``{Neutrino
  Beam Constraints on Flavor-Diagonal Lorentz Violation},''{\em Phys. Rev. D}
  {\bfseries 87} 9, (2013) 096004},
  \href{http://arxiv.org/abs/1302.2598}{{\ttfamily arXiv:1302.2598
  [hep-ph]}}\relax
\mciteBstWouldAddEndPunctfalse
\mciteSetBstMidEndSepPunct{\mcitedefaultmidpunct}
{}{\mcitedefaultseppunct}\relax
\EndOfBibitem
\bibitem{Noordmans:2014bua}
J.~P. Noordmans and K.~K. Vos,
  \href{http://dx.doi.org/10.1103/PhysRevD.89.101702}{``{Limits on Lorentz
  violation from charged-pion decay},''{\em Phys. Rev. D} {\bfseries 89} 10,
  (2014) 101702}, \href{http://arxiv.org/abs/1404.7629}{{\ttfamily
  arXiv:1404.7629 [hep-ph]}}\relax
\mciteBstWouldAddEndPunctfalse
\mciteSetBstMidEndSepPunct{\mcitedefaultmidpunct}
{}{\mcitedefaultseppunct}\relax
\EndOfBibitem
\bibitem{Diaz:2013saa}
J.~S. D\'\i{}az, A.~Kosteleck\'y, and R.~Lehnert,
  \href{http://dx.doi.org/10.1103/PhysRevD.88.071902}{``{Relativity violations
  and beta decay},''{\em Phys. Rev. D} {\bfseries 88} 7, (2013) 071902},
  \href{http://arxiv.org/abs/1305.4636}{{\ttfamily arXiv:1305.4636
  [hep-ph]}}\relax
\mciteBstWouldAddEndPunctfalse
\mciteSetBstMidEndSepPunct{\mcitedefaultmidpunct}
{}{\mcitedefaultseppunct}\relax
\EndOfBibitem
\bibitem{Noordmans:2013dha}
J.~P. Noordmans, H.~W. Wilschut, and R.~G.~E. Timmermans,
  \href{http://dx.doi.org/10.1103/PhysRevLett.111.171601}{``{Limits on Lorentz
  violation from forbidden $\beta$ decays},''{\em Phys. Rev. Lett.} {\bfseries
  111} 17, (2013) 171601}, \href{http://arxiv.org/abs/1308.5570}{{\ttfamily
  arXiv:1308.5570 [hep-ph]}}\relax
\mciteBstWouldAddEndPunctfalse
\mciteSetBstMidEndSepPunct{\mcitedefaultmidpunct}
{}{\mcitedefaultseppunct}\relax
\EndOfBibitem
\bibitem{Fayyazuddin:1994wh}
.~Fayyazuddin and .~Riazuddin, \href{http://dx.doi.org/10.1142/8064}{{\em {A
  Modern Introduction To Particle Physics (3rd Edition)}}}.
\newblock Oxford University Press\relax
\mciteBstWouldAddEndPunctfalse
\mciteSetBstMidEndSepPunct{\mcitedefaultmidpunct}
{}{\mcitedefaultseppunct}\relax
\EndOfBibitem
\bibitem{Barger:1987nn}
V.~D. Barger and R.~J.~N. Phillips,
\newblock {\em {COLLIDER PHYSICS}}\relax
\mciteBstWouldAddEndPunctfalse
\mciteSetBstMidEndSepPunct{\mcitedefaultmidpunct}
{}{\mcitedefaultseppunct}\relax
\EndOfBibitem
\bibitem{Nakanishi:2002sv}
N.~Nakanishi, ``{T*-product and false nonconservation of angular momentum in
  the pion decay},'' \href{http://dx.doi.org/10.1142/S0217732302006217}{{\em
  Mod. Phys. Lett. A} {\bfseries 17} (2002) 89--93}\relax
\mciteBstWouldAddEndPunctfalse
\mciteSetBstMidEndSepPunct{\mcitedefaultmidpunct}
{}{\mcitedefaultseppunct}\relax
\EndOfBibitem
\bibitem{Jacob:1959at}
M.~Jacob and G.~C. Wick, ``{On the General Theory of Collisions for Particles
  with Spin},'' \href{http://dx.doi.org/10.1016/0003-4916(59)90051-X}{{\em
  Annals Phys.} {\bfseries 7} (1959) 404--428}\relax
\mciteBstWouldAddEndPunctfalse
\mciteSetBstMidEndSepPunct{\mcitedefaultmidpunct}
{}{\mcitedefaultseppunct}\relax
\EndOfBibitem
\bibitem{Gibson:1976wp}
W.~M. Gibson and B.~R. Pollard, {\em {Symmetry Principles in Elementary
  Particle Physics}}.
\newblock Cambridge University Press\relax
\mciteBstWouldAddEndPunctfalse
\mciteSetBstMidEndSepPunct{\mcitedefaultmidpunct}
{}{\mcitedefaultseppunct}\relax
\EndOfBibitem
\bibitem{Leader:2011vwq}
E.~Leader, {\em {Spin in particle physics}},
\newblock vol.~15\relax
\mciteBstWouldAddEndPunctfalse
\mciteSetBstMidEndSepPunct{\mcitedefaultmidpunct}
{}{\mcitedefaultseppunct}\relax
\EndOfBibitem
\bibitem{Perkins:2000xb}
D.~H. Perkins, {\em {Introduction to high energy physics}}.
\newblock Cambridge University Press, 4th~ed.\relax
\mciteBstWouldAddEndPunctfalse
\mciteSetBstMidEndSepPunct{\mcitedefaultmidpunct}
{}{\mcitedefaultseppunct}\relax
\EndOfBibitem
\bibitem{UA1:1985bfo}
{UA1}, G.~Arnison {\em et~al.}, ``{Recent Results on Intermediate Vector Boson
  Properties at the CERN Super Proton Synchrotron Collider},''
  \href{http://dx.doi.org/10.1016/0370-2693(86)91603-5}{{\em Phys. Lett. B}
  {\bfseries 166} (1986) 484--490}\relax
\mciteBstWouldAddEndPunctfalse
\mciteSetBstMidEndSepPunct{\mcitedefaultmidpunct}
{}{\mcitedefaultseppunct}\relax
\EndOfBibitem
\bibitem{UA1:1985our}
{UA1}, G.~Arnison {\em et~al.}, ``{Intermediate Vector Boson Properties at the
  {CERN} Super Proton Synchrotron Collider},''
  \href{http://dx.doi.org/10.1209/0295-5075/1/7/002}{{\em EPL} {\bfseries 1}
  (1986) 327--345}\relax
\mciteBstWouldAddEndPunctfalse
\mciteSetBstMidEndSepPunct{\mcitedefaultmidpunct}
{}{\mcitedefaultseppunct}\relax
\EndOfBibitem
\bibitem{UA1:1988rck}
{UA1}, C.~Albajar {\em et~al.}, ``{Studies of Intermediate Vector Boson
  Production and Decay in UA1 at the CERN Proton - Antiproton Collider},''
  \href{http://dx.doi.org/10.1007/BF01548582}{{\em Z. Phys. C} {\bfseries 44}
  (1989) 15--61}\relax
\mciteBstWouldAddEndPunctfalse
\mciteSetBstMidEndSepPunct{\mcitedefaultmidpunct}
{}{\mcitedefaultseppunct}\relax
\EndOfBibitem
\bibitem{D0:2000ydn}
{D0}, B.~Abbott {\em et~al.}, ``{Measurement of the angular distribution of
  electrons from $W \to e \nu$ decays observed in $p\bar{p}$ collisions at
  $\sqrt{s} = 1.8$ TeV},''
  \href{http://dx.doi.org/10.1103/PhysRevD.63.072001}{{\em Phys. Rev. D}
  {\bfseries 63} (2001) 072001},
  \href{http://arxiv.org/abs/hep-ex/0009034}{{\ttfamily
  arXiv:hep-ex/0009034}}\relax
\mciteBstWouldAddEndPunctfalse
\mciteSetBstMidEndSepPunct{\mcitedefaultmidpunct}
{}{\mcitedefaultseppunct}\relax
\EndOfBibitem
\bibitem{CDF:2003tlc}
{CDF}, D.~Acosta {\em et~al.}, ``{Measurement of the polar-angle distribution
  of leptons from $W$ boson decay as a function of the W transverse momentum in
  $p\bar{p}$ collisions at $\sqrt{s} = 1.8$ TeV},''
  \href{http://dx.doi.org/10.1103/PhysRevD.70.032004}{{\em Phys. Rev. D}
  {\bfseries 70} (2004) 032004},
  \href{http://arxiv.org/abs/hep-ex/0311050}{{\ttfamily
  arXiv:hep-ex/0311050}}\relax
\mciteBstWouldAddEndPunctfalse
\mciteSetBstMidEndSepPunct{\mcitedefaultmidpunct}
{}{\mcitedefaultseppunct}\relax
\EndOfBibitem
\bibitem{ATLAS:2011qdp}
{ATLAS}, G.~Aad {\em et~al.}, ``{Measurement of the inclusive $W^\pm$ and
  Z/gamma cross sections in the electron and muon decay channels in $pp$
  collisions at $\sqrt{s}=7$ TeV with the ATLAS detector},''
  \href{http://dx.doi.org/10.1103/PhysRevD.85.072004}{{\em Phys. Rev. D}
  {\bfseries 85} (2012) 072004},
  \href{http://arxiv.org/abs/1109.5141}{{\ttfamily arXiv:1109.5141
  [hep-ex]}}\relax
\mciteBstWouldAddEndPunctfalse
\mciteSetBstMidEndSepPunct{\mcitedefaultmidpunct}
{}{\mcitedefaultseppunct}\relax
\EndOfBibitem
\bibitem{CMS:2020cph}
{CMS}, A.~M. Sirunyan {\em et~al.},
  \href{http://dx.doi.org/10.1103/PhysRevD.102.092012}{``{Measurements of the
  $W$ boson rapidity, helicity, double-differential cross sections, and charge
  asymmetry in $pp$ collisions at $\sqrt {s}$ = 13 TeV},''{\em Phys. Rev. D}
  {\bfseries 102} 9, (2020) 092012},
  \href{http://arxiv.org/abs/2008.04174}{{\ttfamily arXiv:2008.04174
  [hep-ex]}}\relax
\mciteBstWouldAddEndPunctfalse
\mciteSetBstMidEndSepPunct{\mcitedefaultmidpunct}
{}{\mcitedefaultseppunct}\relax
\EndOfBibitem
\bibitem{D0:2013tnv}
{D0}, V.~M. Abazov {\em et~al.}, ``{Evidence for S-Channel Single Top Quark
  Production in $p\bar{p}$ Collisions at $\sqrt{s}$ = 1.96 TeV},''
  \href{http://dx.doi.org/10.1016/j.physletb.2013.09.048}{{\em Phys. Lett. B}
  {\bfseries 726} (2013) 656--664},
  \href{http://arxiv.org/abs/1307.0731}{{\ttfamily arXiv:1307.0731
  [hep-ex]}}\relax
\mciteBstWouldAddEndPunctfalse
\mciteSetBstMidEndSepPunct{\mcitedefaultmidpunct}
{}{\mcitedefaultseppunct}\relax
\EndOfBibitem
\bibitem{CDF:2014uma}
{CDF, D0}, T.~A. Aaltonen {\em et~al.}, ``{Observation of s-channel production
  of single top quarks at the Tevatron},''
  \href{http://dx.doi.org/10.1103/PhysRevLett.112.231803}{{\em Phys. Rev.
  Lett.} {\bfseries 112} (2014) 231803},
  \href{http://arxiv.org/abs/1402.5126}{{\ttfamily arXiv:1402.5126
  [hep-ex]}}\relax
\mciteBstWouldAddEndPunctfalse
\mciteSetBstMidEndSepPunct{\mcitedefaultmidpunct}
{}{\mcitedefaultseppunct}\relax
\EndOfBibitem
\bibitem{ATLAS:2015jmq}
{ATLAS}, G.~Aad {\em et~al.}, ``{Evidence for single top-quark production in
  the $s$-channel in proton-proton collisions at $\sqrt{s}=$8 TeV with the
  ATLAS detector using the Matrix Element Method},''
  \href{http://dx.doi.org/10.1016/j.physletb.2016.03.017}{{\em Phys. Lett. B}
  {\bfseries 756} (2016) 228--246},
  \href{http://arxiv.org/abs/1511.05980}{{\ttfamily arXiv:1511.05980
  [hep-ex]}}\relax
\mciteBstWouldAddEndPunctfalse
\mciteSetBstMidEndSepPunct{\mcitedefaultmidpunct}
{}{\mcitedefaultseppunct}\relax
\EndOfBibitem
\bibitem{CMS:2016xoq}
{CMS}, V.~Khachatryan {\em et~al.}, ``{Search for s channel single top quark
  production in pp collisions at $ \sqrt{s}=7 $ and 8 TeV},''
  \href{http://dx.doi.org/10.1007/JHEP09(2016)027}{{\em JHEP} {\bfseries 09}
  (2016) 027}, \href{http://arxiv.org/abs/1603.02555}{{\ttfamily
  arXiv:1603.02555 [hep-ex]}}\relax
\mciteBstWouldAddEndPunctfalse
\mciteSetBstMidEndSepPunct{\mcitedefaultmidpunct}
{}{\mcitedefaultseppunct}\relax
\EndOfBibitem
\end{mcitethebibliography}

\end{document}